\documentclass{article}
\usepackage{spconf,amsmath,graphicx}
\usepackage{amssymb}
\usepackage{array}

\title{Feature-Based Image Clustering and Segmentation Using Wavelets}
\name{Junyu Chen, Eric C. Frey}
\address{Johns Hopkins University\\
Department of Electrical and Computer Engineering, Whiting School of Engineering
}
\begin{document}

\maketitle
\begin{abstract}
Pixel intensity is a widely used feature for clustering and segmentation algorithms, the resulting segmentation using only intensity values might suffer from noises and lack of spatial context information. Wavelet transform is often used for image denoising and classification. We proposed a novel method to incorporate Wavelet features in segmentation and clustering algorithms. The conventional K-means, Fuzzy c-means (FCM), and Active contour without edges (ACWE) algorithms were modified to adapt Wavelet features, leading to robust clustering/segmentation algorithms. A weighting parameter to control the weight of low-frequency sub-band information was also introduced. The new algorithms showed the capability to converge to different segmentation results based on the frequency information derived from the Wavelet sub-bands.
\end{abstract}
\section{Introduction}
\label{sec:intro}
Many real images are often corrupted by noise in their acquisition, such as biomedical images like Single-photon emission computed tomography (SPECT) images, and the image itself often contains both smooth and textured regions. Wavelet decomposition is a heavily used tool in image processing and classification; it enables the decomposition of an image into varies frequency sub-bands, similar to the way the human visual system operates \cite{Huang:4586427}. Because of this property, Wavelets has been widely adopted in image denoising and texture classification \cite{Chang2000}\cite{Elad2006}. However, there has been limited work on image clustering and segmentation with Wavelets coefficients. 

\begin{figure}[htb]
  \centering
  \centerline{\includegraphics[width=7.5cm]{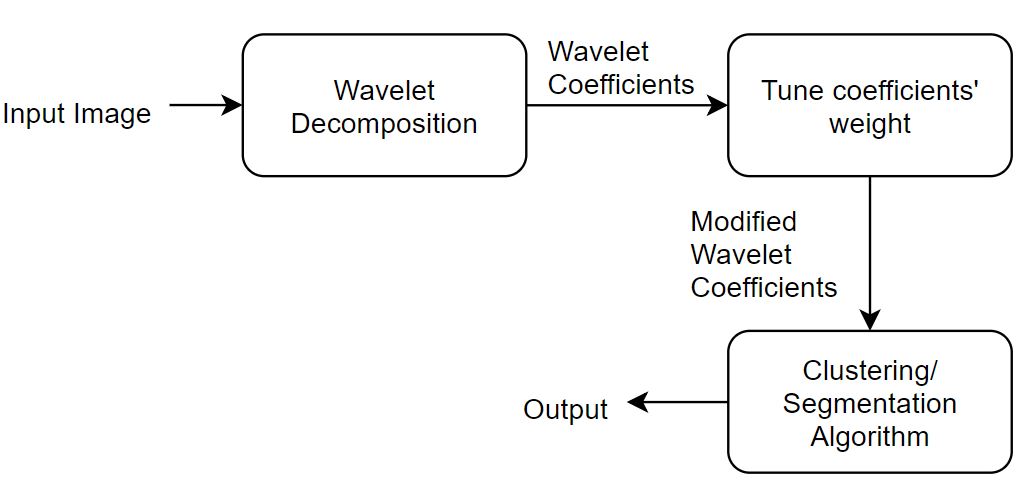}}
  \caption{Overview of algorithmic pipeline.}
\label{fig:pipeline}
\end{figure}

Most conventional image clustering or segmentation algorithms, such as K-means, Fuzzy C-means (FCM), Gaussian mixture model (GMM), and Active contour without edges (ACWE), are based only on image intensities. Some improvements in the clustering algorithms to incorporate information about spatial context were proposed \cite{Pham2001}\cite{Jha2010}\cite{Krinidis:2010}, these methods usually require some modifications of the objective functions, thus they complicate calculations and increases computational time. A Wavelet based image clustering scheme was introduced by Porter and Canagarajah \cite{Porter:491343}, the authors proposed a K-means image segmentation algorithm using the optimal Wavelet features derived from the image. In their method, the algorithm assigns pixels into two categories, smooth regions and textured regions by a certain threshold in the Wavelet domain, then K-means algorithm was applied in the different categories. In this paper, we propose to apply clustering or segmentation algorithms directly on the Wavelet domain with a weighting parameter to the Wavelet coefficients. The weighting parameter provides the possibility to control the different information to be segmented, the schematic of the algorithm is shown in Fig. \ref{fig:pipeline}. Three algorithms, K-means\_w, FCM\_w, and ACWE\_w are proposed, and they are the modified versions of the conventional K-means, FCM, and ACWE algorithms to incorporate Wavelet coefficients. By the nature of the Wavelet transform, the spatial information is automatically being incorporated in the Wavelet coefficients, so the penalty terms for incorporating the spatial context \cite{Pham2001}\cite{Jha2010}\cite{Krinidis:2010} are not necessary. The detailed methods and implementation will be described in section \ref{sec:theory}.

\section{Image Clustering and Segmentation in Wavelet Domain}
\label{sec:theory}
\subsection{Wavelet Decomposition}
\begin{figure}[htb]
  \centering
  \centerline{\includegraphics[width=9.5cm]{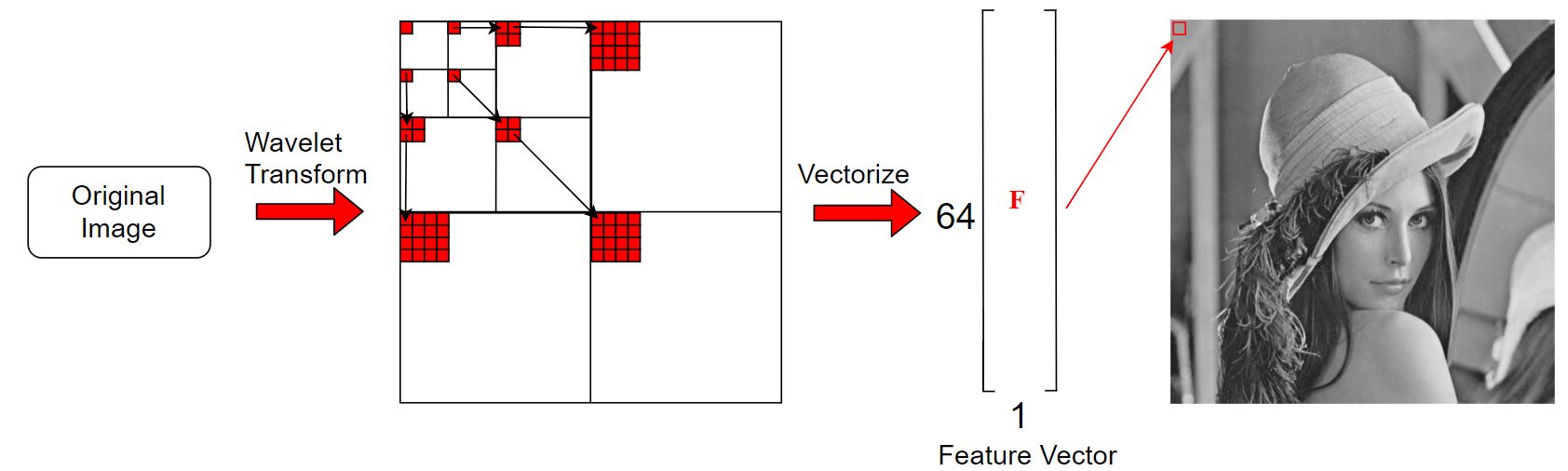}}
  \caption{Global Wavelet transform.}
\label{fig:globaltrans}
\end{figure}

The wavelet transform represents the singularity content of an image at multiple scales \cite{Choi2001}. In our wavelet decomposition scheme, as shown in Fig. \ref{fig:globaltrans}, the transform is applied on the entire image and the first Wavelet tree was extracted and then vectorized. The vectorized Wavelet tree represents the feature vector of the first pixel. We then shift the image by one pixel and apply the transform again, iteratively.

\begin{figure}[htb]
  \centering
  \centerline{\includegraphics[width=2.5cm]{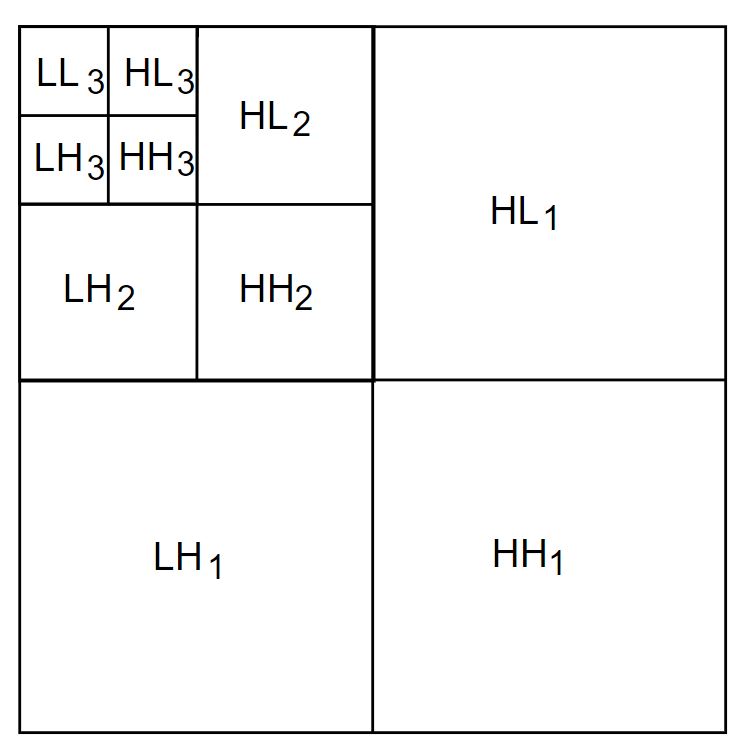}}
  \caption{Sub-bands of 2-D Wavelet transform.}
\label{fig:wavedec}
\end{figure}

Fig. \ref{fig:wavedec} shows the labeled sub-bands of a 3-level transform, the sub-bands $HL_k$, $LH_k$, and $HH_k$ are called the details, where $k=1,2,...,K$, $K$ represents the maximum decomposition level; the sub-band $LL_K$ is called the smooth approximation; the sub-bands, $LL_3$, $HL_3$, $LH_3$, and $HH_3$, are recognized as low-frequency bands \cite{Porter:491343}. We then introduced a weighting parameter $w$ to the low-frequency bands in the vectorized Wavelet feature vector, the vector is defined as:
$$F_i = \left( \begin{smallmatrix} \text{vec}(LL_3)^w\\ \text{vec}(HL_3)^w \\ \text{vec}(LH_3)^w \\ \text{vec}(HH_3)^w \\ \text{vec}(HL_2) \\ \text{vec}(LH_2) \\ ...\end{smallmatrix} \right),$$
\noindent where the $F_i$ represents the feature vector of $i^{th}$ pixel, $\text{vec}()$ is the vectorization operator, and weighting parameter $w$ biases the differences in low-frequency bands when computing the squared differences in the clustering/segmentation algorithms, the effect of this parameter will be discussed in the next sub-sections.

\subsection{K-means using Wavelet feature vectors}
The K-means clustering algorithm aims to minimize the squared distances between all pixel intensity and the cluster center\cite{Tou:103883}. The algorithm iteratively minimizes the following objective function\cite{Bankman:2000}:
\begin{equation}
	\label{eq:1} J_{K-means} = \sum_{j \in \Omega}\sum_{k=1}^{C}z_{jk}\Vert y_j-v_k\Vert^2,
\end{equation} 

\noindent where $\Omega$ is the image domain, $C$ is the number of classes, $y_j$ is the observation at pixel j, and $v_k$ is the centroid of class $k$. 

The objective function of proposed K-means\_w is define as:
\begin{equation}
	\label{eq:2} J_{K-means\_w} = \sum_{j \in \Omega}\sum_{k=1}^{C}z_{jk}\Vert F_j-\bar{F}_k\Vert^2,
\end{equation} 
\noindent where $F_j$ is the feature vector at pixel $j$, the term $\bar{F}_k$ is the centroid of feature vectors of class $k$, and the hard membership function $z$ is defined as:
\begin{equation}
	\label{eq:3} z_{jk} = 
	\begin{cases}
      1, & \text{if}\ \text{voxel $j$ is in class $k$}\\
      0, & \text{otherwise}
    \end{cases},
\end{equation} 
\noindent and
\begin{equation}
	\label{eq:4} \sum_{k=1}^{C}z_{jk} = 1, \ \ \forall j.
\end{equation}

The algorithm computes the squared error between the feature vector at each pixel and the mean feature vector of each cluster; this operation spread the energy to each sub-bands evenly. The weighting parameter $w$, which we introduced previously, controls the weight of the low-frequency sub-bands. With $w>1$, the squared error in low-frequency sub-bands gets emphasized, on the other hand, the squared error in low-frequency sub-bands gets suppressed when $w<1$. 

\subsection{Fuzzy c-means using Wavelet feature vectors}
The FCM clustering algorithm was proposed by Bezdek, it is an improvement of the hard membership K-means clustering algorithm\cite{BEZDEK1984191}. With the similar idea as K-means, The algorithm assigns soft membership of each class to a given pixel, depending on the similarity of the pixel intensity value to a particular class relative to all other classes. In the FCM\_w, pixel intensities were replaced by the Wavelet feature vectors:
\begin{equation}
	\label{eq:5} J_{FCM\_w} = \sum_{j \in \Omega}\sum_{k=1}^{C}u_{jk}^q\Vert F_j-\bar{F}_k\Vert^2,
\end{equation} 

\noindent the following constrains remains the same as the conventional FCM:
\begin{equation}
   \sum_{k=1}^{C}u_{jk}=1, \ \ \forall j.
\end{equation}
\noindent where $\Omega$ is the image domain, $C$ is the number of classes, $u$ represents the soft membership function, $F_j$ is the feature vector at pixel $j$, $\bar{F}_k$ is the centroid of feature vectors of class $k$, and parameter $q$ is a weighting exponent that satisfies $q>1$, it controls the amount of fuzzy overlap between clusters\cite{BEZDEK1984191}. The larger $q$ values indicate a greater degree of overlap (and vice versa). By controlling $w$ in feature vectors $F$, we control the importance of the low-frequency bands.

Applying the Lagrange multiplier method to minimize the objective function $J_{FCM}$, an iterative update expression of membership functions and cluster centroids can be obtained.

\subsection{Active contour without edges using Wavelet feature vectors}
ACWE was developed by Chan and Vese\cite{Chan:902291}, the algorithm is based on the techniques of curve evolution, the Mumford-Shah functional for segmentation, and level sets\cite{Morar:6356188}\cite{OSHER198812}\cite{mackay_1997}. The Chan-Vese function is defined as:
\begin{multline}\label{eq:6}
   J_{ACWE}(c_1,c_2,\phi)=\lambda_1\int_\Omega(f-c_1)^2H(\phi)dx\\ 
   +\lambda_2\int_\Omega(f-c_2)^2(1-H(\phi))dx +\mu\int_\Omega\vert\bigtriangledown H(\phi)\vert dx,
\end{multline}
\noindent where $\phi$ denotes the level set function. The parameters $\lambda_1$ and $\lambda_2$ are fixed to be $\lambda$ (i.e. $\lambda_1=\lambda_2=\lambda $), $\mu$ is a scale parameter for determining the weight of the length term. The term $H(\phi)$ denotes the Heaviside function of $\phi$ defined as\cite{Chan:902291}:

\[
  H(z)=\begin{cases}
               1 & \text{if}\ z\geqslant 1 \\
               0 & \text{if}\ z < 1, \\
            \end{cases}
\]
\noindent it represents the foreground and background regions in image $f$, and $\int_\Omega\vert\bigtriangledown H(\phi)\vert$ represents the length of the contour of the region. $c_1$ and $c_2$ denote the mean intensities of foreground and background regions defined by $\phi \geqslant 0$ and $\phi < 0$. The segmentation of a region can be obtained by minimizing $J_{ACWE}(c_1,c_2,\phi)$ respect to $c_1$, $c_2$ and $\phi$. 
To apply ACWE in the wavelet domain, the energy function (i.e., $J_{ACWE}$) was modified, and the objective function of the ACWE\_w is:
\begin{multline}\label{eq:7}
   J_{ACWE\_w}(\bar{F}_1,\bar{F}_2,\phi)=\lambda_1\int_\Omega(F-\bar{F}_1)^2H(\phi)dx\\ 
   +\lambda_2\int_\Omega(F-\bar{F}_2)^2(1-H(\phi))dx +\mu\int_\Omega\vert\bigtriangledown H(\phi)\vert dx,
\end{multline}
\noindent where $F$ is the Wavelet feature vector, $\bar{F}_1$ and $\bar{F}_2$ indicate the mean Wavelet feature vectors inside foreground and background regions, respectively. Again, by adjusting the weighting parameter, $w$, we decide how much information in the low-frequency bands we choose to incorporate.

\section{Results}
\label{sec:result}
\begin{figure}[htb]
  \centering
  \centerline{\includegraphics[width=6cm]{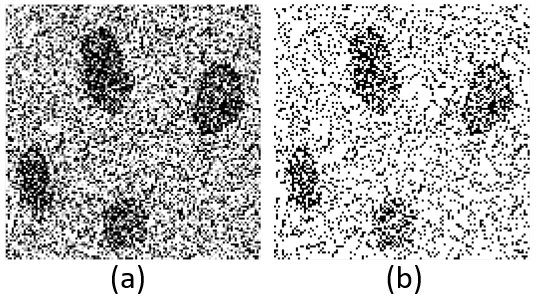}}
  \caption{(a) contains simulated noisy minefield image and (b) contains binarized minefield image using Otsu's method.}
\label{fig:Mine}
\end{figure}

\begin{figure}[htb]
  \centering
  \centerline{\includegraphics[width=8.5cm]{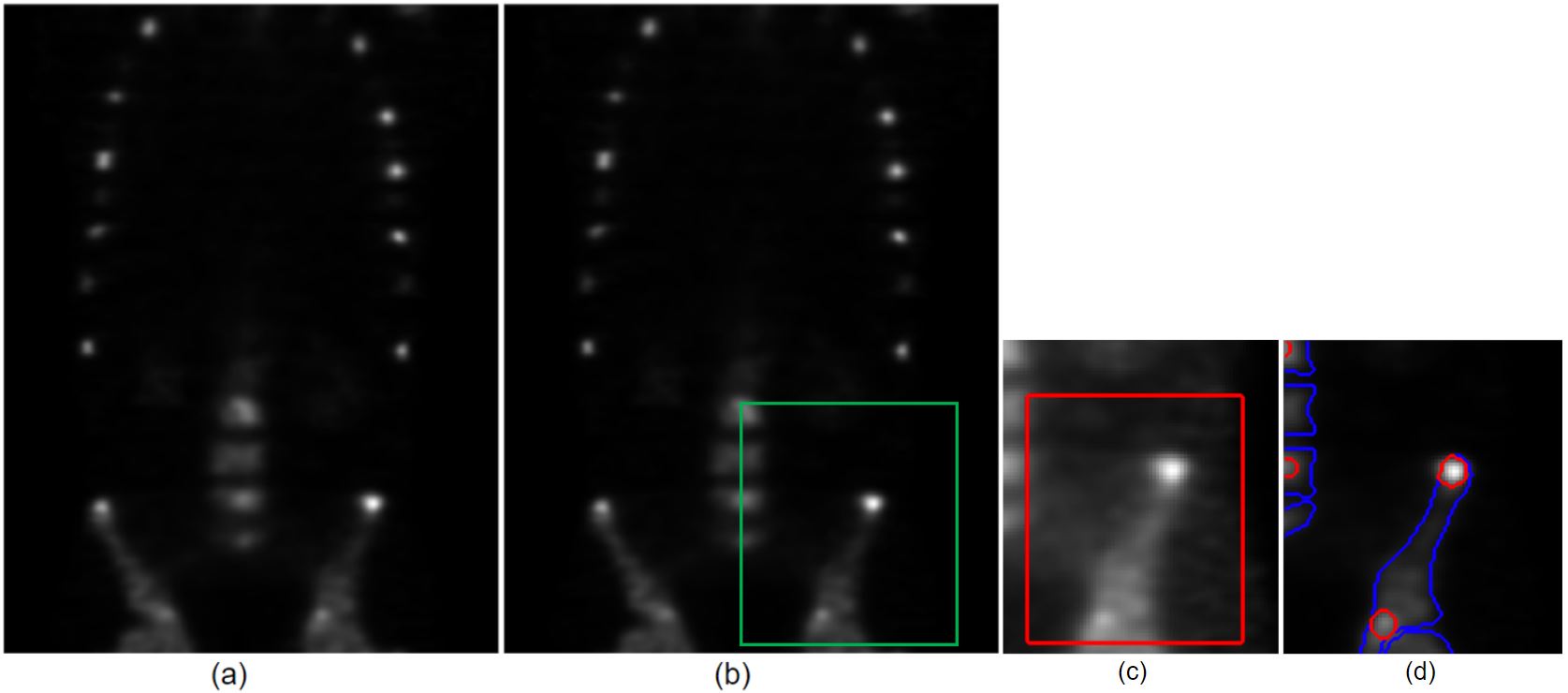}}
  \caption{Image (a) contains a slice of QBSPECT image, where image (b) shows the QBSPECT image with rectangular region of interest, image (c) shows the cropped region with initial contour of ACWE (red curve), and (d) shows the truth bone regions (blue curves) and lesion regions (red curves).}
\label{fig:SPECT}
\end{figure}

The proposed K-means and FCM algorithms were tested on binarized minefield image, as shown in Fig. \ref{fig:Mine};
the proposed ACWE-W algorithm was tested on the simulated Quantitative Bone SPECT image using the NURBS-based XCAT phantom \cite{Chen2019}, an example slice is shown in Fig. \ref{fig:SPECT} (a). In Fig. \ref{fig:SPECT} (b), the green curve indicates the rectangular region of interest that we cropped, Fig. \ref{fig:SPECT} (c) represents the cropped image with the initial contour (red curve) of ACWE, and Fig. \ref{fig:SPECT} (c) contains the ground truth of lesion regions (red curves) and bone regions (blue curves).

The effect of using different orthogonal and bi-orthogonal Wavelet filter pairs was also studied. The filter coefficients are shown in Table. \ref{table:filter}.

\begin{figure}[htb]
  \centering
  \includegraphics[width=3cm]{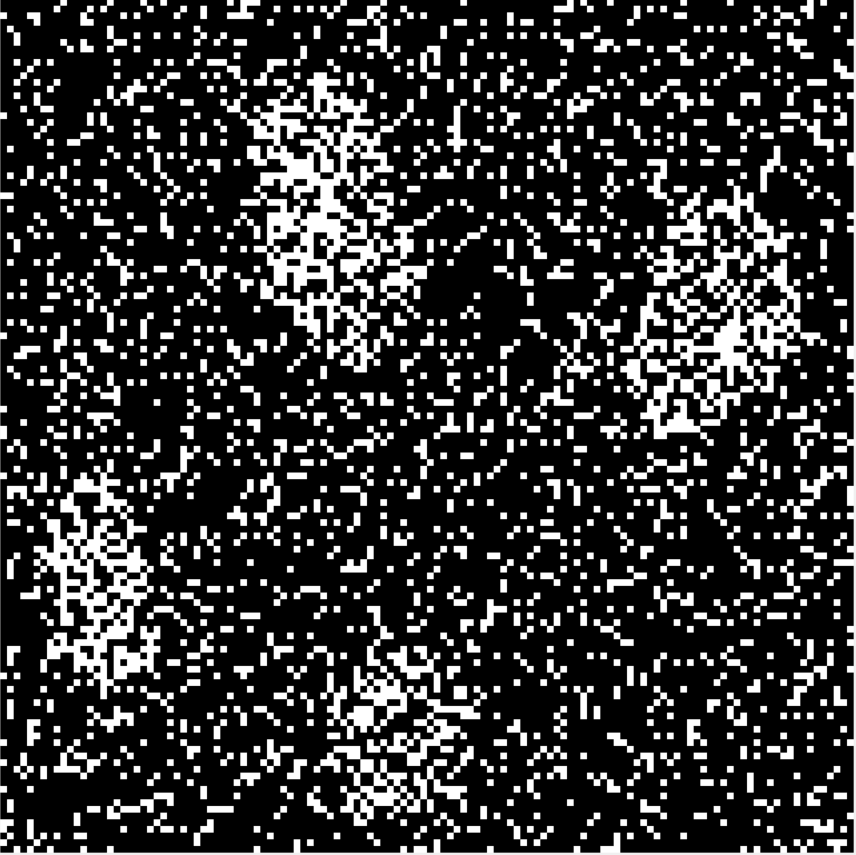}
  \caption{Figure shows the clustering result generated by conventional K-means and FCM.}
\label{fig:fcm_km}
\end{figure}

\begin{table}[h]
\caption{Bi-orthogonal and Orthogonal filter pairs}
\label{table:filter}
\begin{center}
\begin{tabular}{||c | m{18em}||} 
 \hline
 8/8 & Orthogonal pair ('$o_1$')\\
 \hline
 $h_0$ & $[0.2304,0.7148, 0.6309,-0.0280,$ $-0.1870, 0.0308, -0.0329, -0.0106]$ \\
 \hline
 $f_0$ & $[-0.0106,0.0329, 0.0308, -0.1870,$ $-0.0280, 0.6309, 0.7148, 0.2304]$ \\ [1ex] 
 \hline\hline
 6/10 & Bi-orthogonal pair ('$bio_1$') \\ [0.5ex] 
 \hline
 $h_0$ & $[-0.1291,0.0477, 0.7885, 0.7885,$ $0.0477, -0.1291]$ \\
 \hline
 $f_0$ & $[0.0189, 0.0070, -0.0672, 0.1334, 0.6151,$ $0.6151, 0.1334, -0.0672, 0.0070, 0.0189]$ \\
 \hline\hline
 9/7 & Bi-orthogonal pair ('$bio_2$') \\ [0.5ex] 
 \hline
 $h_0$ & $[-0.0161,-0.0424, 0.0680, 0.3960,$ $0.6033, 0.3960,0.0680,-0.0424,-0.0161]$ \\
 \hline
 $f_0$ & $[0.1513, -0.3980, 0.2022, 1.5032, 0.2022,$ $-0.3980, 0.1513]$ \\
 \hline
\end{tabular}
\end{center}
\end{table}

\begin{figure}[htb]
  \centering
  \centerline{\includegraphics[width=7.5cm]{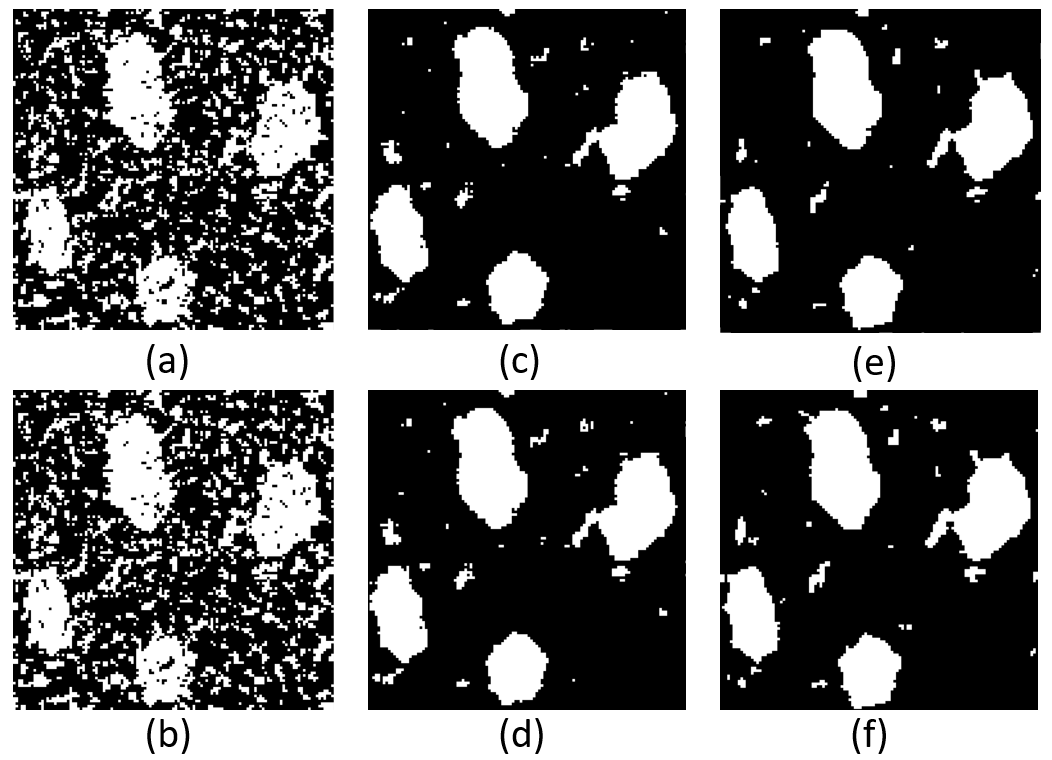}}
  \caption{Results generated by proposed K-means algorithm. The first column generated with $w=0.7$, middle column generated with $w=1$, the third column generated with $w=3$.}
\label{fig:km_fcm_dwt}
\end{figure}

The results obtained by the conventional 2-class K-means and FCM algorithm are the same, as shown in Fig. \ref{fig:fcm_km}. The clustering results were corrupted by noises because the algorithms depend only on local intensity values and spatial information was missing. Fig. \ref{fig:km_fcm_dwt} shows the results generated by the proposed K-means\_w and FCM\_w algorithms. The first column in Fig. \ref{fig:km_fcm_dwt} represents the results produced by K-means\_w with the weight parameter, $w = 2$; the second column shows the results produced by FCM\_w; the first row were generated with orthogonal filter pairs, $o_1$, shown in the Table \ref{table:filter}; the second and the third row were generated with bi-orthogonal filter pairs, $bio_1$ and $bio_2$, respectively. With $w = 2$, the importance of high-frequency information (i.e., noises) gets suppressed, and the algorithms depend more on low-frequency information. Thus, clustering algorithms produce smoother results. The results also suggest that the use of bi-orthogonal filter pairs in K-means\_w and FCM\_w algorithms could lead to a better clustering result.

\begin{figure}[htb]
  \centering
  \centerline{\includegraphics[width=3cm]{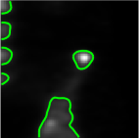}}
  \caption{Segmentation generated by conventional ACWE algorithm.}
\label{fig:acwe_o}
\end{figure}

\begin{figure}[htb]
  \centering
  \centerline{\includegraphics[width=7.5cm]{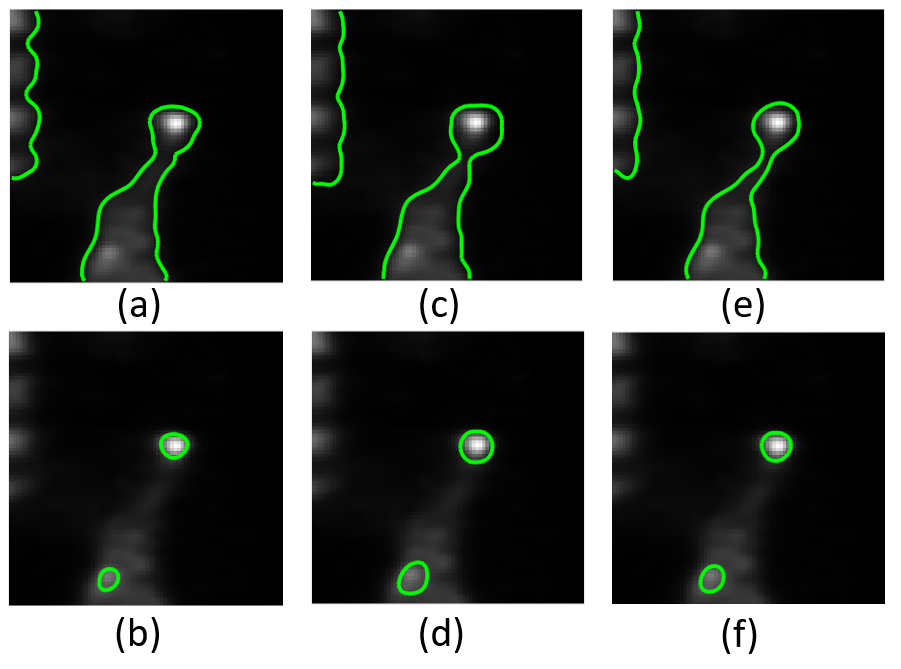}}
  \caption{Segmentation results generated by ACWE\_w algorithm. The row was generated by $o_1$ filter pairs with $w=0.75$ and $2$, middle row was generated by $bio_1$ with $w=0.7$ and $2$, the third row was generated by $bio_2$ with $w=0.8$ and $2$.}
\label{fig:acwe_dwt}
\end{figure}

Fig. \ref{fig:acwe_o} shows the results produced by conventional ACWE algorithm. Comparing to the ground truth (Fig. \ref{fig:SPECT} (d)), the algorithm failed to obtain a good segmentation for either bone or lesion. The results generated by the ACWE\_w algorithm are shown in Fig. \ref{fig:acwe_dwt}. When $w<1$, the algorithm incorporates high-frequency patterns to segment the bone regions. In contrast, when $w>1$, the algorithm focuses more on the low-frequency patterns, which provides the ability to segment lesion regions. The results also suggest that for the ACWE\_w algorithm, the differences in the resulting segmentation is negligible with different Wavelet filter pairs.

\section{Conclusion}
\label{sec:Conclusion}
Image clustering and segmentation algorithms often concentrate only on local pixel intensities, although many algorithms have been proposed to incorporate spatial context \cite{Pham2001}\cite{Jha2010}\cite{Krinidis:2010}, the modified clustering/segmentation algorithms could still be sensitive to noises. In this paper, we proposed to apply clustering/segmentation algorithms in the Wavelet domain, and we introduced a weighting parameter to control the importance of the low-frequency sub-bands. The results provided that the proposed methods perform robust to noises, and the algorithms converge to different meaningful regions based on different frequency band. 

\section{ACKNOWLEDGMENT}
The authors would like to thank Shuwen Wei who provided expertise that greatly assisted this work.

\newpage
\bibliographystyle{IEEEbib}

\end{document}